\documentclass[
reprint,
amsmath,amssymb,
aps,
prl,
longbibliography,
superscriptaddress,
]{revtex4-2}

\usepackage{graphicx}
\usepackage{dcolumn}
\usepackage{bm}
\usepackage{bbold}
\usepackage{hyperref}
\usepackage{amsmath}
\usepackage{amsthm}

\usepackage{braket}
\makeatletter

\newcommand{\Tr}{\text{Tr}}

\begin{document}

\preprint{APS/123-QED}

\title{Universal Sensitivity Bound for Thermal Quantum Dynamic Sensing}

\author{Rui Zhang}
\thanks{These authors contributed equally to this work.}
\affiliation{Zhejiang Key Laboratory of Quantum State Control and Optical Field Manipulation, Department of Physics, Zhejiang Sci-Tech University, 310018 Zhejiang, China}
\affiliation{Department of Physics, Zhejiang University, 310018 Zhejiang, China}

\author{Yang Yang}
\thanks{These authors contributed equally to this work.}
\affiliation{Zhejiang Key Laboratory of Quantum State Control and Optical Field Manipulation, Department of Physics, Zhejiang Sci-Tech University, 310018 Zhejiang, China}
\affiliation{Department of Physics, Zhejiang University, 310018 Zhejiang, China}

\author{Wenkui Ding}
\email{wenkuiding@zstu.edu.cn}
\affiliation{Zhejiang Key Laboratory of Quantum State Control and Optical Field Manipulation, Department of Physics, Zhejiang Sci-Tech University, 310018 Zhejiang, China}

\author{Xiaoguang Wang}
\email{xgwang@zstu.edu.cn}
\affiliation{Zhejiang Key Laboratory of Quantum State Control and Optical Field Manipulation, Department of Physics, Zhejiang Sci-Tech University, 310018 Zhejiang, China}

\begin{abstract}

This work unifies the equilibrium and non-equilibrium frameworks of quantum metrology within the context of many-body systems. We investigate dynamic sensing schemes to derive an upper bound on the quantum Fisher information for probe states in thermal equilibrium with their environment.
We establish that the dynamic quantum Fisher information for a thermal probe state is upper bounded by the degree of non-commutation between the transformed local generator and the Hamiltonian for the thermal state.
Furthermore, we show that this upper bound scales as the square of the product of the inverse temperature and the evolution time.
In the low-temperature limit, we establish an additional upper bound expressed as the seminorm of the commutator divided by the energy gap.
We apply this thermal dynamic sensing scheme to various models, demonstrating that the dynamic quantum Fisher information satisfies the established upper bounds.

\end{abstract}

\date{\today}
\maketitle


\textit{Introduction---}
Quantum metrology aims to achieve optimal sensitivity with given quantum resources~\cite{braunstein1994statistical,giovannetti2006quantum,giovannetti2011advances,degen2017quantum}. The most widely researched area involves utilizing entangled states as probes in Ramsey interferometry to reach the Heisenberg limit~\cite{pezze2018quantum}.
However, the fragility of entangled quantum states makes quantum-enhanced sensitivity difficult to achieve in realistic experimental implementations~\cite{braun2018quantum,escher2011general,demkowicz2012elusive}. Interactions within the quantum many-body system and coupling to the environment lead to decoherence and relaxation, causing entangled states to rapidly decay into classical or thermal equilibrium states~\cite{deutsch2018eigenstate,montenegro2025quantum}.
Conversely, it is possible to achieve quantum-enhanced sensitivity by appropriately harnessing interactions within a quantum many-body system~\cite{mishra2021driving}. For instance, these interactions can induce quantum phase transitions, leading to various proposals for realizing quantum critical metrology~\cite{rams2018at,garbe2020critical}.

Since thermal states are easy to prepare and robust against decoherence, quantum metrology using these states has recently attracted significant attention~\cite{hauke2016measuring,mehboudi2016achieving,miller2018energy,meng2025bounds,tumbiolo2025shake}. Specifically, for quantum parameter estimation, we can employ the Gibbs state $\rho = \frac{e^{-\beta H_\lambda}}{Z}$, where $Z$ is the partition function and $H_\lambda$ is the Hamiltonian dependent on the parameter $\lambda$. The inverse temperature is defined as $\beta = 1/(k_B T)$, where $k_B$ is the Boltzmann constant and $T$ is the environmental temperature. While estimating $\beta$ corresponds to quantum thermometry~\cite{correa2015individual,mehboudi2019thermometry,mehboudi2022fundamental}, the more general goal is to estimate the parameter $\lambda$ encoded in the thermal state.
Generally, the Hamiltonian $H_\lambda$ incorporates non-linear terms to facilitate quantum-enhanced sensitivity\cite{beau2017nonlinear,sensitivity2019li}. The interplay between linear and non-linear interactions induces a quantum phase transition~\cite{zanardi2007bures}. In the low-temperature limit, the thermal state $\rho$ reduces to the ground state $|\Psi_0(\lambda)\rangle$, aligning the parameter estimation process with the ground-state fidelity approach in criticality-enhanced metrology~\cite{frerot2018quantum,quantum2008zanardi,criticality2021wu}. Here, the quantum Fisher information and fidelity susceptibility exhibit divergent behavior near the critical point, indicating a significant enhancement in sensitivity.

However, the ground-state overlap scheme for criticality metrology faces a crucial problem known as critical slowing down: adiabatic driving near the critical point requires a significantly long time due to the vanishing energy gap. When this evolution time is taken into account, the enhanced sensitivity is often diminished~\cite{rams2018at}. A similar issue persists in equilibrium thermal sensing schemes, where time dependence is often ill-defined. In this work, we investigate non-equilibrium thermal sensing, where the quantum Fisher information exhibits explicit time dependence. Specifically, we use a thermal state as the probe and encode the parameter via unitary time evolution~\cite{dynamic2021chu}. This non-equilibrium scheme aligns more closely with experimental implementations while retaining the advantage of using easily prepared thermal states.

In quantum metrology and parameter estimation, the objective is to estimate an unknown parameter $\lambda$ from a parameter-dependent state $\rho_\lambda$. The ultimate precision of this estimation is governed by the quantum Cramér-Rao bound, $\delta \lambda\geq 1/F_\lambda$, where $F_\lambda$ is the quantum Fisher information (QFI), a central quantity in metrology~\cite{braunstein1994statistical}. The QFI is defined as $F_\lambda=\text{Tr}[\rho_\lambda L^2]$, where $L$ is the symmetric logarithmic derivative (SLD) satisfying $\partial \rho_\lambda/\partial\lambda=\frac{1}{2}(\rho_\lambda L+L\rho_\lambda)$.For a general encoding process governed by a unitary evolution operator $U_\lambda$, the parameter is estimated from the evolved state $\rho_\lambda=U_\lambda \rho_0 U_\lambda^\dagger$, where $\rho_0$ is a parameter-independent probe state~\cite{de2013quantum}. When $\rho_0$ is spectrally decomposed as $\rho_0=\sum_{i=1}^M p_i|\psi_i\rangle\langle\psi_i|$ (where $M$ is the dimension of the support), the QFI can be expressed in terms of the eigenvalues $p_i$ and eigenstates $|\psi_i\rangle$~\cite{quantum2014liu,liu2015quantum}:
\begin{equation}
\label{eq:dynamic_qfi_mixed}
F_\lambda=\sum_{i=1}^M4p_i \text{Var}[h_\lambda]|_{|\psi_i\rangle}-\sum_{i\neq j}\frac{8p_ip_j}{p_i+p_j}|\langle\psi_i|h_\lambda|\psi_j\rangle|^2,
\end{equation}
where $h_\lambda=iU_\lambda^\dagger\frac{\partial U_\lambda}{\partial\lambda}$ is the transformed local generator, a crucial quantity in dynamic sensing schemes~\cite{pang2014quantum}. The variance is defined as $\text{Var}[h_\lambda]\big|_{|\psi_i\rangle}=\langle\psi_i|h_\lambda^2|\psi_i\rangle-\langle\psi_i|h_\lambda|\psi_i\rangle^2$. Notably, if the probe is a pure state $\rho_0=|\Psi_0\rangle\langle\Psi_0|$, the QFI simplifies to $F_\lambda=4(\langle\Psi_0|h_\lambda^2|\Psi_0\rangle-\langle\Psi_0|h_\lambda|\Psi_0\rangle^2)$.

\begin{figure}
\includegraphics[width=0.5\textwidth]{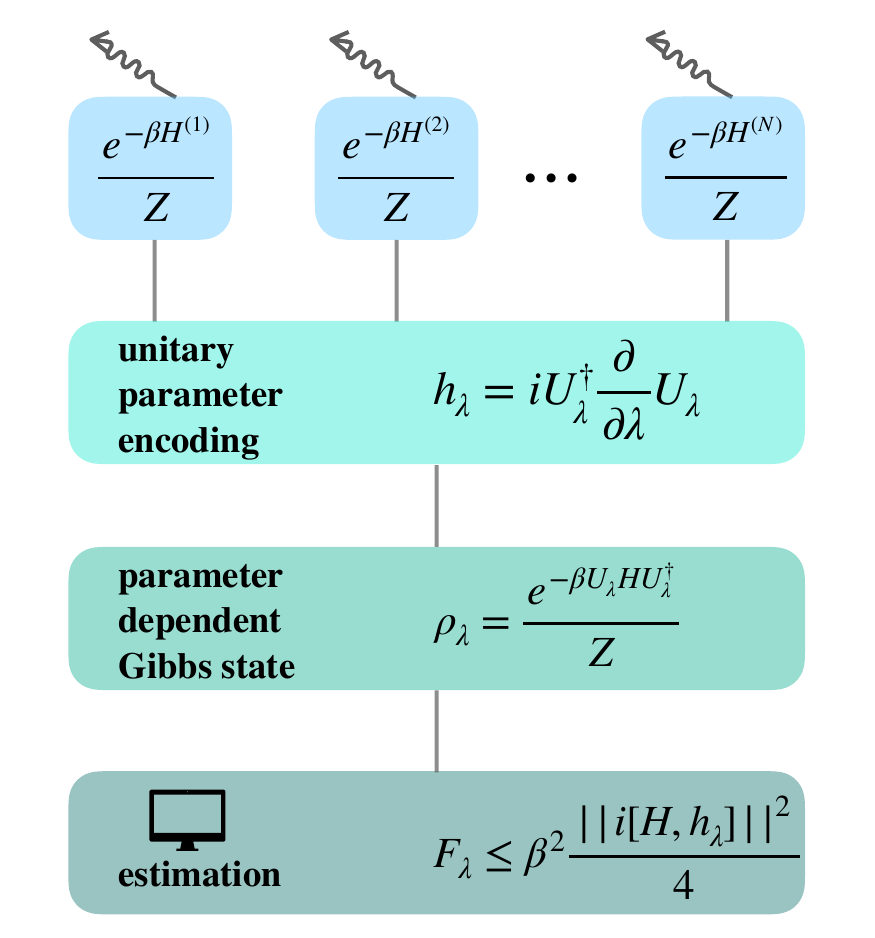}
\centering
\caption{\label{fig:sketch} Schematics of the thermal dynamic sensing scheme. Gibbs state, $\rho_0=e^{-\beta H}/Z$ with $H=\sum_{k=1}^N H^{(k)}$, is used as the probe state and the parameter encoding process $U_\lambda$ is unitary. The parameter-dependent state $\rho_\lambda$ is still a Gibbs state, and we can obtain a general upper bound for the quantum Fisher information which is determined by the seminorm of the commutator $||i[H,h_\lambda]||$, where the transformed local generator $h_\lambda$ naturally appears.}
\end{figure}

\textit{General upper bound for dynamic QFI with thermal probe states---}
We consider a scenario where the initial state is the Gibbs state, $\rho_0=e^{-\beta H}/Z$, where $\beta=1/(k_B T)$ is the inverse temperature and $Z=\text{Tr}[e^{-\beta H}]$ is the partition function~\cite{troiani2018quantum}.
The parameter $\lambda$ is encoded into the quantum state via the unitary evolution $U_\lambda$, yielding the evolved parameter-dependent state $\rho_\lambda=U_\lambda \rho_0 U_\lambda^\dagger$.
Since the unitary transformation commutes with the exponential function, $\rho_\lambda$ is equivalent to a thermal state with a transformed Hamiltonian: $\rho_\lambda={e^{-\beta U_\lambda H U_\lambda^\dagger}}/{Z}$.
Crucially, the partition function $Z$ remains independent of the parameter $\lambda$.
If the Hamiltonian $H$ is spectrally decomposed as $H=\sum_i E_i|\psi_i\rangle\langle\psi_i|$, the QFI can be calculated explicitly as follows:
\begin{equation}
\label{eq:F_dynamic_general}
\begin{aligned}
F_\lambda=&\beta^2\text{Var}[i[H,h_\lambda]]|_{\rho_0}\\
&-\frac{\beta^2}{Z}\sum_{i\neq j}e^{-\beta E_i}[1-\text{tanhc}^2(\frac{\beta\Delta_{ij}}{2})]|\langle\psi_i|i[H,h_\lambda]|\psi_j\rangle|^2,
\end{aligned}
\end{equation}
where the variance of an operator $\hat{A}$ with respect to a quantum state $\rho$ is defined as $\text{Var}[\hat{A}]|_{\rho}=\text{Tr}[\rho\hat{A}^2]-(\text{Tr}[\rho \hat{A}])^2$.
The transformed local generator is $h_\lambda=iU_\lambda^\dagger \frac{\partial U_\lambda}{\partial\lambda}$, and $\Delta_{ij}=E_i-E_j$.
The cardinal hyperbolic tangent function is defined as $\text{tanhc}(x)=\frac{\tanh(x)}{x}$, and $0\leq \text{tanhc}^2(x) \leq 1$.

The second term in the expression for $F_\lambda$ is non-negative (since $0\leq \text{tanhc}^2(x) \leq 1$ and $e^{-\beta E_i}\geq 0$ ).
This observation immediately leads to the following universal upper bound:
\begin{equation}
\label{eq:dynamic_thermal_bound}
F_\lambda\leq \beta^2\text{Var}[i[H,h_\lambda]]|_{\rho_0}\leq \beta^2\frac{||i[H,h_\lambda]||^2}{4}.
\end{equation}
The second inequality uses the property that the variance of any operator is bounded by its seminorm~\cite{boixo2007generalized}, $\text{Var}[\hat{A}]|_\rho\leq ||A||^2/4$.
The seminorm is defined as the difference between the largest eigenvalue and smallest eigenvalue, $||A||=E_\text{max}-E_\text{min}$.
This inequality represents a key theoretical result, indicating that the QFI for thermal metrology is bounded by the inverse temperature ($\beta$) and the spectral width of the commutator ($||i[H,h_\lambda]||$).
Compared to bounds found in Ref.~\cite{fundamental2025abiuso,garcia2024estimation}, the Hamiltonian derivative $\frac{\partial H_\lambda}{\partial\lambda}$ is replaced by the hermitian operator $i[H,h_\lambda]$, and the variance is specifically taken with respect to the initial probe state $\rho_0$.
This inequality immediately reveals two necessary physical conditions for extracting information about the parameter $\lambda$ using a thermal probe state.
First, in the high-temperature limit, the explicit $\beta^2$ dependence in the upper bound, indicates a vanishing QFI as $\beta\rightarrow 0$.
This is physically sound because, in the high-temperature limit, the probe state $\rho_0$ approaches the identity matrix (maximally mixed state).
The system essentially ceases to evolve meaningfully, preventing any information from being encoded into the quantum state.
Second, assuming a finite Hilbert space dimension, the non-commutativity between the Hamiltonian $H$ and the generator $h_\lambda$ can be quantified by the measure $c_{H,h_\lambda}=||i[H,h_\lambda]||$.
The inequality shows that larger non-commutativity corresponds to a higher upper bound on the QFI, highlighting that non-commutating terms are essential for achieving enhanced sensitivity in dynamic thermal sensing schemes. 

Furthermore, for any two hermitian operators $\hat{A}$ and $\hat{B}$, the inequality $||i[\hat{A},\hat{B}]||\leq ||\hat{A}||\cdot||\hat{B}||$ holds.
When the unitary time evolution is governed by a time-independent Hamiltonian $U_\lambda(t)=e^{i\tilde{H}_\lambda t}$, we can derive the following upper bound:
\begin{equation}
\label{eq:dynamic_thermal_bound_2}
F_\lambda\leq \beta^2\frac{||i[H,h_\lambda]||^2}{4}\leq\beta^2 t^2\frac{||H||^2||\frac{\partial\tilde{H}_\lambda}{\partial\lambda}||^2}{4}.
\end{equation}
Here, we utilized the property that  the seminorm of the transformed local generator is bounded by $||h_\lambda||\leq t||\frac{\partial\tilde{H}_\lambda}{\partial\lambda}||$.
This stems from the integral representation $h_\lambda=\int_0^t U_\lambda^\dagger (s)\frac{\partial \tilde{H}_\lambda}{\partial\lambda}U_\lambda(s)ds$.
Since unitary transformations preserve the operator spectrum, applying the triangle inequality yields the linear scaling with $t$~\cite{fundamental2023ding}.
In practice, the seminorm of the Hamiltonian derivative $\frac{\partial\tilde{H}_\lambda}{\partial\lambda}$ is significantly easier to evaluate than that of $h_\lambda$.
Crucially, this final upper bound exhibits explicit simultaneous scaling with $\beta$ and $t$, which is essential for characterizing time-dependent QFI.

The upper bounds derived previously exhibit a dependence on temperature scaling as $\propto \beta^2$.
Consequently, these bounds become trivial (divergent) in the low-temperature limit where $\beta\rightarrow \infty$.
To address this, we derive an alternative upper bound that remains useful in this regime:
\begin{equation}
F_\lambda \leq \sum_{i=1} 4p_i\text{Var}[ h_\lambda]_{|\psi_i\rangle}\leq 4\frac{\text{Var}[i[H,h_\lambda]]|_{\rho_0}}{\Delta^2}\leq \frac{||i[H,h_\lambda]||^2}{\Delta^2},
\end{equation}
where $\Delta=\min_{i\neq j}\left|\Delta_{ij}\right|$ represents the minimum non-zero energy gap of the Hamiltonian $H$.
The first inequality follows directly from Eq.~(\ref{eq:dynamic_qfi_mixed}) and reflects the convexity of the QFI.
The second inequality explicitly captures the dependence on the minimal energy gap $\Delta$. Therefore, this bound is particularly relevant for analyzing dynamic sensing scheme near critical points in quantum many-body systems, where the energy gap closes.

\textit{Examples---}We begin by examining a concrete example to illustrate these upper bounds.
Consider a spin-$J$ system ($J\geq 1$) initialized in the thermal state $\rho_0=e^{-\beta J_z}/Z$, where $J_\alpha$ denotes the spin angular momentum operators.
The unitary time evolution is governed by $U_\lambda=e^{-it\lambda J_\alpha}$, leading to the transformed local generator $h_\lambda=t J_\alpha$.
Consequently, the commutator in the bound becomes $i[H,h_\lambda]=i[J_z,tJ_\alpha]$.
If $\alpha=z$, the commutator vanishes, implying no information can be obtained.
However, if $\alpha=x$, we first calculate the seminorm bound.
Using the commutation relation $[J_z,J_x]=iJ_y$, we find $F_\lambda\leq \beta^2 t^2 ||J_y||^2/4$.
Assuming $J=N/2$, this scale as $\propto N^2$, suggesting Heisenberg scaling at finite temperatures.
Next, we analyze the tighter bound provided by the variance.
For $\alpha=x$, the relevant quantity is $\text{Var}[i[H,h_\lambda]]|_{\rho_0}=t^2\text{Var}[J_y]|_{\rho_0}$.
Explicitly, $\text{Var}[J_y]|_{\rho_0}=\Tr[\rho_0 J_y^2]=\frac{1}{2}(J(J+1)-\frac{Z_2}{Z})$, where $Z=\sum_{M=-J}^J \exp(-\beta M)$ and $Z_2=\sum_{M=-J}^J \exp(-\beta M)M^2$.
This variance bound has an exact analytic form:
\begin{equation}
\begin{aligned}
\beta^2\text{Var}&[i[H,h_\lambda]]|_{\rho_0}=\\
-\beta^2&\left[\frac{1}{8} \sinh (\beta ) \text{csch}^3\left(\frac{\beta }{2}\right) ((J+1) \sinh (\beta  J)\right.\\
&\left.-J \sinh (\beta  (J+1))) \text{csch}\left(\beta  \left(J+\frac{1}{2}\right)\right)\right].
\end{aligned}
\end{equation}
In the large-spin limit ($J\gg 1$), this variance asymptotically approaches $\text{Var}[J_y]|_{\rho_0}\approx \frac{1}{4}(2J+1)\coth{\frac{\beta}{2}}-\frac{1}{4}\coth^2\frac{\beta}{2}$.
This result scales linearly with $J$, indicating that the sensitivity is actually limited to the standard quantum limit.
Thus, the variance bound is significantly tighter than the semi-norm bound.
Finally, calculating the exact QFI for this scheme yields:
\begin{equation}
F_\lambda=2t^2\tanh{\frac{\beta}{2}}[(J+\frac{1}{2})\coth(\beta(J+\frac{1}{2}))-\frac{1}{2}\coth{\frac{\beta}{2}}].
\end{equation}
As expected, $\lim_{\beta\rightarrow 0}F_\lambda=0$, confirming no sensitivity in the high-temperature limit.
For $J\gg 1$, the exact QFI approximates to $F_\lambda\approx 2J-2(2J+1)/(1+e^\beta)$.
This confirms that as $\beta\rightarrow\infty$, $F_\lambda\propto 2J$, signifying the standard quantum limit.

Subsequently, we consider a thermal state as the probe for a non-linear metrology scheme.
The initial state of the sensing protocol is given by the thermal state $\rho_0=\frac{e^{-\beta J_z}}{Z}$.
The parameter encoding is governed by the unitary $U_\lambda=e^{-i\lambda J_x^2 t}$, which describes a one-axis twisting process~\cite{ma2011quantum}.
In this scenario, the transformed local generator is calculated as $h_\lambda=iU_\lambda^\dagger \frac{\partial}{\partial\lambda}U_\lambda=t J_x^2$.
Applying the formula in Eq.~(\ref{eq:F_dynamic_general}), we determine the QFI.
The matrix element $|\langle JM|h_\lambda|JM^\prime\rangle|^2$ is non-zero only when $M^\prime=M$ or $M^\prime=M\pm 2$.
Specifically, for $M^\prime=M$, we obtain $|\langle JM|h_\lambda|JM^\prime\rangle|^2=\frac{1}{4}[J(J+1)-M^2]^2$. 
Conversely, for $M^\prime=M\pm 2$, the term becomes $|\langle JM|h_\lambda|JM^\prime\rangle|^2=\frac{1}{16}(J\mp M)(J\mp M-1)(J\pm M+1)(J\pm M+2)$.
Consequently, the QFI is derived as:
\begin{equation}
\begin{aligned}
F_\lambda=\frac{t^2}{2}\coth^2(\frac{\beta}{2})\text{sech}(\beta)\eta,
\end{aligned}
\end{equation}
with
\begin{equation*}
\begin{aligned}
&\eta=3-4J(J+1)+\text{csch}[\beta(J+\frac{1}{2})]\times\\
&\{J(2J-1)\sinh[\beta(J+\frac{3}{2})]+(J+1)(2J+3)\sinh[\beta(J-\frac{1}{2})]\}
\end{aligned}
\end{equation*}
Next, the variance-based upper bound is calculated as:
\begin{equation}
\beta^2\text{Var}[i[H,h_\lambda]]|_{\rho_0}=\frac{1}{8}\beta^2 t^2\cosh(\beta)\text{csch}^4(\frac{\beta}{2})\eta.
\end{equation}
To determine the upper bound in terms of the seminorm, we evaluate the seminorm of the commutator.
We have $i[H,h_\lambda]=it[J_z,J_x^2]=-t(J_xJ_y+J_yJ_x)$.
Recall that the seminorm of an operator is defined as the difference between its maximal and minimal eigenvalues.
As the exact eigenvalues of $J_xJ_y+J_yJ_x$ are difficult to obtain analytically, we employ a semi-classical treatment.
Since unitary transformations preserve eigenvalues, we define a rotated operator $\hat{C}=e^{i\frac{\pi}{4}J_z}(J_xJ_y+J_yJ_x)e^{-i\frac{\pi}{4}J_z}=J_x^2-J_y^2$.
The seminorm of the commutator is therefore equivalent to the seminorm of $\hat{C}$.
In the classical limit ($J\gg 1$), the angular momentum vector $\mathbf{J}$ has fixed length $J$, parameterized by $J_x=J\sin\theta\cos\phi$, $J_y=J\sin\theta\sin\phi$ and $J_z=J\cos\theta$.
Thus, $\hat{C}\approx J^2\sin^2\theta\cos{2\phi}$.
The maximum value, $\hat{C}_\text{max}\approx J^2$, occurs when $\sin\theta=1$ and $\cos(2\phi)=1$.
Similarly, the minimum value is $-J^2$, yielding a classical range $\hat{C}\in[-J^2,J^2]$.
For large $J$, quantum fluctuations are negligible relative to $J^2$, allowing the eigenvalues of $\hat{C}$ to approximate these classical values. 
Consequently, the seminorm of $\hat{C}$ is approximately $2J^2$.
In conclusion, for $J\gg 1$, the seminorm of the commutator is $\approx 2J^2$, leading to $\beta^2\frac{||i[H,h_\lambda]||^2}{4}\approx\beta^2 t^2 J^4$.
This scaling corresponds to the ultimate sensitivity limit of non-linear metrology  using $\lambda J_x^2$ as the parameter encoding Hamiltonian (see Ref.~\cite{imai2025metrological}).
Note that this ultimate limit requires an entangled initial state, whereas the thermal state is limited to a scaling $\propto J^2$, as derived above.
Finally, the upper bound in terms of the separate seminorms is given by $\beta^2\frac{||H||^2||h_\lambda||^2}{4}=\beta^2t^2\frac{||J_z||^2||J_x^2||^2}{4}=\beta^2t^2 J^6$.
\begin{figure}
\includegraphics[width=0.5\textwidth]{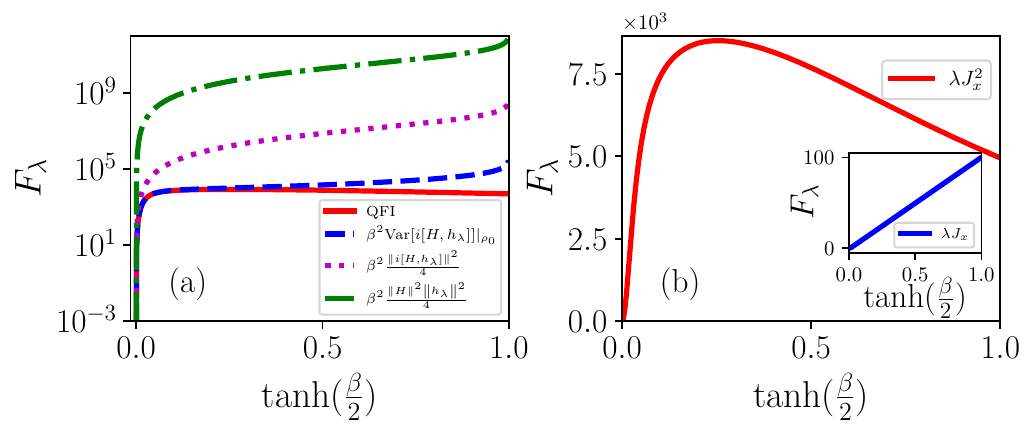}
\centering
\caption{\label{fig:Jx2} QFI as a function of $P=\tanh{\frac{\beta}{2}}$ for the thermal state $\rho_0=e^{-\beta J_z}/Z$. (a) QFI and its upper bound for a non-linear parameter encoding process $U_\lambda=e^{-i\lambda t J_x^2}$. (b) Comparison of the dynamic QFI of the non-linear parameter encoding process $U_\lambda=e^{-i\lambda t J_x^2}$ and the linear parameter encoding process $U_\lambda=e^{-i\lambda t J_x}$.}
\end{figure}

In Fig.~\ref{fig:Jx2}(a), we plot the QFI and the corresponding upper bounds as a function of the polarization $P=\tanh{({\beta}/{2})}$ for the non-linear parameter encoding process defined by $U_\lambda=e^{-i\lambda J_x^2 t}$.
There results numerically verify the validity of the derived inequality.
In Fig.~\ref{fig:Jx2}(b), we compare the QFI of the linear encoding scheme, $U_\lambda=e^{-i\lambda J_x t}$ , with that of the non-linear scheme.
The comparison reveals distinct thermal behaviors: the non-linear parameter encoding process exhibits a specific optimal temperature $\beta$ that maximizes the QFI.
In contrast, the linear encoding shows a monotonic dependence on temperature, where lower temperatures consistently yield superior sensitivity.

\begin{figure}
\includegraphics[width=0.5\textwidth]{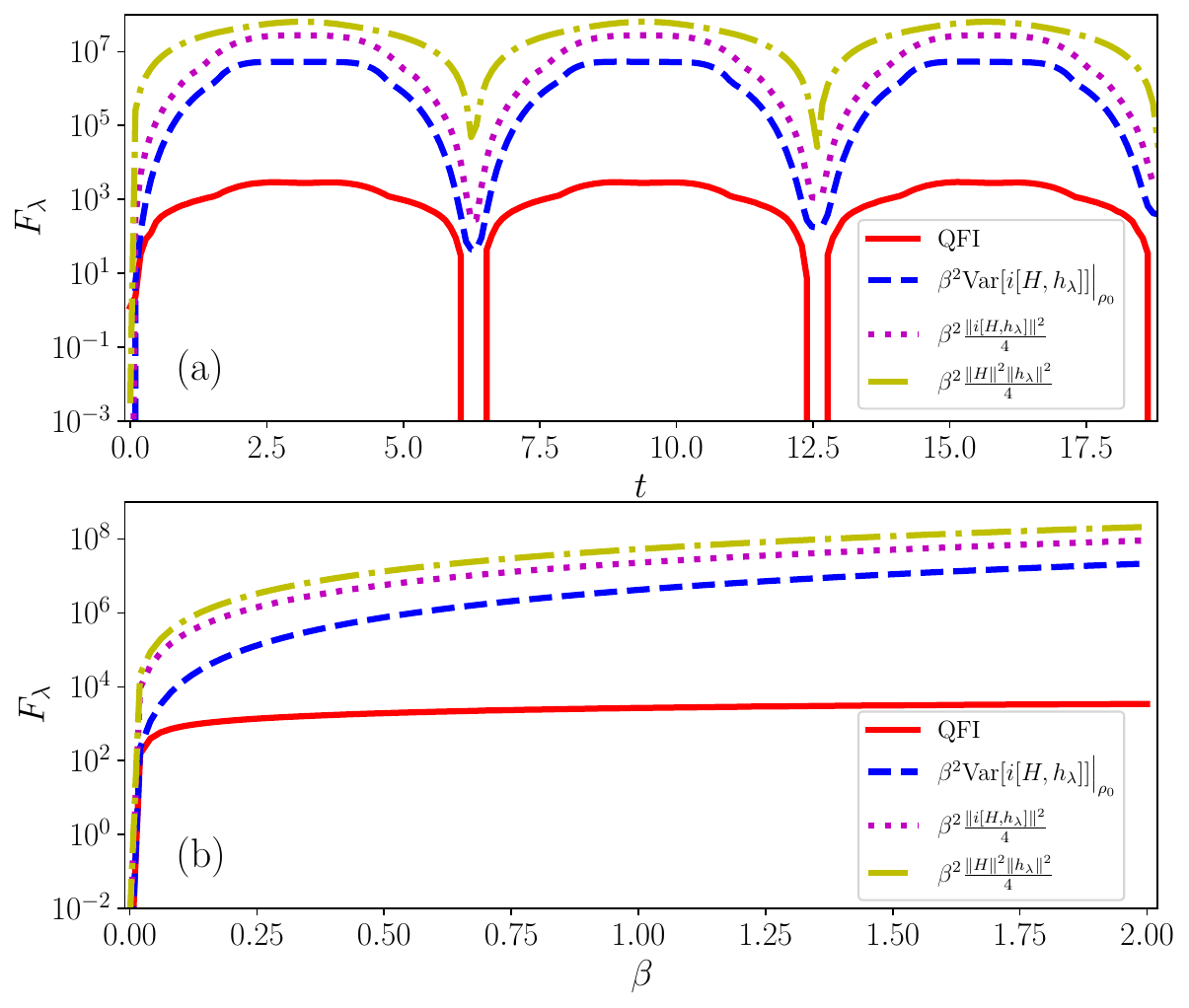}
\centering
\caption{\label{fig:example_LMG_1}QFI for the LMG model. (a) QFI and its bound as a function of evolution time, for a fixed inverse temperature $\beta=1.1$. (b) QFI and its bound as a function of inverse temperature $\beta$, for a fixed evolution time $t=3.14$.}
\end{figure}

In the end, we consider a more general scenario where the parameter encoding process is governed  by the Lipkin-Meshkov-Glick (LMG) Hamiltonian~\cite{salvatori2014quantum}, defined as $H_{\text{LMG}}=J_x^2+\lambda J_z$.
The probe state remains the Gibbs state $\rho_0=e^{-\beta J_z}/Z$.
In this case, the transformed local generator is given by the integral, $h_\lambda=\int_0^t e^{iH_\lambda t^\prime}\frac{\partial H_\lambda}{\partial\lambda} e^{-iH_\lambda t^\prime}d t^\prime=\int_0^t e^{i(J_x^2+\lambda J_z)t^\prime}J_z e^{-i(J_x^2+\lambda J_z)t^\prime}d t^\prime$.
In Fig.~\ref{fig:example_LMG_1}, we illustrate the resulting QFI alongside the corresponding bounds determined by the derived inequality.

\textit{Discussion---}
In this theoretical work, we investigate a dynamic sensing scheme utilizing thermal states as probes. We establish a universal upper bound on the dynamic QFI, identifying the necessary conditions for achieving quantum-enhanced sensitivity. Beyond the single-parameter estimation focused on here, this framework can be extended to multi-parameter estimation. For instance, one parameter can be encoded directly into the thermal probe state, while another is encoded via time evolution. Specifically, if the probe state is $\rho_0=e^{-\beta H_\theta}/Z$ and the time evolution is $U_\lambda$, it is possible to simultaneously estimate both $\theta$ and $\lambda$. 

Utilizing thermal states has significant practical value, as quantum systems naturally reside in thermal equilibrium with their environment. Conventional sensing schemes typically require an initial step to polarize qubits from a thermal state; achieving quantum-enhanced sensitivity often requires the further generation of entangled states. This process consumes significant time and resources, imposing stringent requirements on the physical implementation. In contrast, thermal states can be used directly or actively prepared with reduced initialization time. Finally, while this work focuses on quantum parameter estimation, similar techniques can be applied to the Hamiltonian learning problem using thermal inputs.

This work was supported by Quantum Science and Technology-National Science and Technology Major Project (Grant No. 2024ZD0301000), the National Natural Science Foundation of China (Grant No. 12305031), the Hangzhou Joint Fund of the Zhejiang Provincial Natural Science Foundation of China under Grant No.
LHZSD24A050001, the Science Foundation of Zhejiang Sci-Tech University (Grants No. 23062088-Y, and No. 23062153-Y).


%

\end{document}